\documentclass[preprint,5p,times,sort&compress,]{elsarticle}
\usepackage[utf8]{inputenc}
\usepackage{footmisc}
\usepackage{lineno,hyperref}
\usepackage{listings}
\usepackage{amssymb}
\usepackage{color}
\usepackage{multirow}
\modulolinenumbers[1]
\usepackage{changes}

\definecolor{maroon}{rgb}{0.5,0,0}
\definecolor{darkgreen}{rgb}{0,0.5,0}
\lstdefinelanguage{XML}
{
  basicstyle=\ttfamily,
  morestring=[s]{"}{"},
  morecomment=[s]{?}{?},
  morecomment=[s]{!--}{--},
  commentstyle=\color{darkgreen},
  moredelim=[s][\color{black}]{>}{<},
  moredelim=[s][\color{red}]{\ }{=},
  stringstyle=\color{blue},
  identifierstyle=\color{maroon}
}
\begin{document}
\begin{frontmatter}

\title{The Astrophysical Multimessenger Observatory Network (AMON): Performance and Science Program.}

\author[psu,igcpg]{Hugo A. Ayala Solares\corref{mycorrespondingauthor}}
\cortext[mycorrespondingauthor]{Corresponding author}

\author[psu,igcpg,astropsu]{Stephane Coutu}

\author[psu,igcpg,astropsu]{D. F. Cowen}

\author[psu,igcpg]{James J. DeLaunay}

\author[igcpg,astropsu,igctoc]{Derek B. Fox}

\author[psu,columbia,columbia2]{Azadeh Keivani}

\author[psu,igcpg,astropsu]{Miguel Mostaf\'a}

\author[psu,igcpg,astropsu]{Kohta Murase}

\author[psu,eso]{Foteini Oikonomou}

\author[irfu]{Monica Seglar-Arroyo}

\author[psu,sp]{Gordana Te\v si\' c}

\author[psu,igcpg]{Colin F. Turley}

\address[psu]{Department of Physics, Pennsylvania State University, University Park, PA 16802, USA}
\address[igcpg]{Center for Particle \& Gravitational Astrophysics, Institute for Gravitation and the Cosmos, Pennsylvania State University, University Park, PA 16802, USA}
\address[astropsu]{Department of Astronomy \& Astrophysics, Pennsylvania State University, University Park, PA 16802, USA}
\address[igctoc]{Center for Theoretical \& Observational Cosmology, Institute for Gravitation and the Cosmos, Pennsylvania State University, University Park, PA 16802, USA}
\address[columbia]{Department of Physics, Columbia University, New York, NY 10027}
\address[columbia2]{Columbia Astrophysics Laboratory, Columbia University, New York, NY 10027, USA}
\address[eso]{European Southern Observatory, Karl-Schwarzschild-Str. 2, D-85748, Garching bei München, Germany}
\address[sp]{Shopify Inc., 150 Elgin Street, Ottawa, ON K2P 1L4, Canada}
\address[irfu]{IRFU, CEA, Université Paris-Saclay, F-91191 Gif-sur-Yvette, France}




\begin{abstract}

The Astrophysical Multimessenger Observatory Network (AMON) has been built with the purpose of enabling near real-time coincidence searches using data from leading multimessenger observatories and astronomical facilities. Its mission is to evoke discovery of multimessenger astrophysical sources, exploit these sources for purposes of astrophysics and fundamental physics, and explore multimessenger datasets for evidence of multimessenger source population
AMON aims to promote the advancement of multimessenger astrophysics by allowing its participants to study the most energetic phenomena in the universe and to help answer some of the outstanding enigmas in astrophysics, fundamental physics, and cosmology. 
The main strength of AMON is its ability to combine and analyze sub-threshold data from different facilities. Such data cannot generally be used stand-alone to identify astrophysical sources. 
The analyses algorithms used by AMON can identify statistically significant coincidence candidates of multimessenger events, leading to the distribution of AMON alerts used by partner observatories for real-time follow-up that may identify and, potentially, confirm the reality of the multimessenger association. We present the science motivation, partner observatories, implementation and summary of the current status of the AMON project.
\end{abstract}

\begin{keyword}
Multimessenger, high-energy astrophysics, gravitational waves, neutrinos, cosmic rays, gamma rays
\end{keyword}

\end{frontmatter}

%\linenumbers

%%%%%%%%%%%%%%%%%
\section{Introduction}

The development and improvement of novel detection methods for the messengers of the four fundamental forces of nature is driving the era of multimessenger astrophysics. 
Each messenger ---neutrino ($\nu$), cosmic ray (CR), photon ($\gamma$) and gravitational wave (GW)--- provides distinct and valuable information of the most violent phenomena in the universe. 
Combining the information from the different messengers will enable us to answer fundamental questions in high-energy astrophysics. 

The first detection of neutrinos, cosmic rays and multiwavelength electromagnetic (EM) radiation were made in the 20\textsuperscript{th} century. Multimessenger observations started with the observation of solar neutrinos and, in 1987, with the observation of coincident neutrinos and photons from supernova SN1987A~\cite{sn1987a}. This joint observation led to an understanding of how stars e\-volve in their last moments, giving insight into physical models of core-collapse supernovae and the formation of neutron stars. Other examples of follow-up observations accomplished at the intersection of neutrinos and EM radiation are the multiwavelength campaign to search for counterparts of an IceCube multiplet~\cite{ic2017} or \textit{Swift} searches for transient sources of high-energy neutrinos~\cite{icrc2017}.
In one of these campaigns, the high-energy IceCube neutrino event \mbox{IceCube-170922A} correlated with a gam\-ma-ray flaring episode detected by the \textit{Fermi}-LAT and MAGIC observatories of the blazar \mbox{TXS 0506+056}, with a significance of association ${>}3\sigma$ ~\cite{nugammaTXS}. This prompted an archival search for high-energy neutrinos prior to 2017 finding that \mbox{TXS 0506+056} is, at 3.5$\sigma$ significance, a possible source of neutrinos~\cite{ICFlare}.
The joint work between the neutrino and EM observatories~\citep[e.g.,][]{Keivani:2018rnh} shows the power of a multimessenger approach in astrophysics and highlights some of the advantages of the Astrophysical Multimessenger Observatory Network (AMON). 

In 2015, the first detection of a GW by the Laser Interferometer Gravitational-Wave Observatory (LIGO) of a coalescence from the coalescence of a compact binary black hole system marked the beginning of GW astronomy~\cite{firstGW}. Two years later, the detection of the first GW detection of a binary neutron star merger heralded the era of multimessenger astrophysics since this event was the first to be observed in both GW and EM channels~\cite{gw170817}. It provided dramatic insight into how mergers of compact objects produce short gamma-ray bursts as well as associated afterglow and kilonova emission.

The majority of current joint analyses, mostly follow-ups, focus on events that strongly exceeded the detection thresholds of each of the observatories, leaving out potential signal events, or ``sub-threshold'' events,  that are difficult to distinguish from background processes (See for example \cite{ic2017,icrc2017} or follow-up observations that appear in the Astronomer's Telegram\footnote{\url{http://www.astronomerstelegram.org/}}).

Combining sub-threshold events from different observatories through coincidence analyses can enhance the signal-to-noise ratio and lead to the discovery of new multimessenger sources.  As a case in point, the high energy neutrino \newline\mbox{IceCube-170922A} detected by IceCube had a 50\% probability of being astrophysical in isolation, but when observed in coincidence with a flaring gamma-ray signal observed by \textit{Fermi} from the same direction and time, became consistent at more than $3\sigma$ with having been emitted by the blazar \mbox{TXS 0506+056}.

AMON is a virtual observatory primary designed to receive and integrate sub-threshold events from several observatories.  As discussed in~\cite{amon2013, amonicrc}, AMON has three main goals:
\begin{itemize}
\item Perform coincidence searches of sub-threshold events of different observatories in real-time, and distribute prompt alerts to follow-up observatories. 
\item Receive events and broadcast them, through the Gamma-Ray Coordinates Network/Transient Astronomy Network (GCN/TAN)\footnote{\url{https://gcn.gsfc.nasa.gov}}~\cite{gcn}, to the astronomical community for follow-up.
\item Store events into its database to perform archival coincidence searches.
\end{itemize}
AMON is a network\footnote{Any collaboration, group or scientist interested in joining AMON, are encouraged to contact the authors for more information.} that can be expanded for new triggering and follow-up facilities through a memorandum of understanding (MoU)\footnote{\url{http://www.amon.psu.edu/join-amon/}\label{mou}}. 

In this paper, we present the features of AMON, as well as the ongoing status of the project.
In Section~\ref{sec:science} we present the scientific justification for the project. Section~\ref{sec:frame} we describe the AMON framework, infrastructure and software. 
We define the different AMON multimessenger channels for follow-up in Section~\ref{sec:channels}, as well as a tested procedure that can be followed to do coincidence analyses. We present current results in Section~\ref{sec:results} and we conclude in Section~\ref{sec:concl}.

%%%%%%%%%%%%%%%%%%%%%%%%%%%%%%%%%%%%%%%%%%%%%%%%
%%%%%%%%%%%%%%% SCIENCE %%%%%%%%%%%%%%%%%%%%%%%%
%%%%%%%%%%%%%%%%%%%%%%%%%%%%%%%%%%%%%%%%%%%%%%%%
\section{Science Cases}\label{sec:science}

AMON provides a framework to search for astrophysical multimessenger sources where physical processes occur under the most extreme conditions. 
We will review some of the most compelling science cases.

{\it Cosmic particle accelerators}: AMON seek to identify the sources and accelerators of highly-energetic cosmic rays, one of the main open questions in the science of high-energy astrophysics~\cite{originCR}.  Since the directions of neutrinos and gamma-rays are not changed by magnetic fields when traveling through the universe, observations of these messengers can shed light on the sources and the acceleration processes responsible for the production of cosmic rays. 
The IceCube observatory was able to detect an astrophysical flux of neutrinos at the level of ${\sim}3\times10^{-8}$\,GeV\,cm$^{-2}$\,s$^{-1}$\,sr$^{-1}$~\cite{icnature,firstpevnu,Aartsen:2013jdh,cosmicnu}. 
The neutrino flux in the 10-100 TeV range may be even higher~\cite{Aartsen:2014muf,Aartsen:2015knd}, which suggests that high-energy neutrinos are produced in dense environments and they have an important contribution to the energy budget of non-thermal particles~\cite{Murase:2015xka}. 
No individual sources of high-energy neutrinos has been established, although the blazar TXS 0506+056 is a candidate with a ${>}3\sigma$ evidence.
With the advent of the new generation of neutrino detectors, including IceCube-Gen2 and KM3Net, it will be more likely to find more sources, including the bulk origin of the diffuse neutrino flux, especially with the combination of EM observations in X-rays and gamma-rays~\cite{kmwaxman}. In addition, coincidence analyses between gamma-ray and neutrino channels are critical for the search of the sources of ultra-high-energy cosmic rays (UHECRs)~\cite{hillas}. Such searches can test the unification of production mechanisms of neutrinos, gamma rays and UHECRs~\cite{kmwaxman,Fang:2017zjf}.

{\it Ultra-high-energy cosmic ray accelerators}: UHECR detectors such as the Pierre Auger Observatory have sensitivity to charged cosmic rays, photons and neutrons with energy exceeding ${\sim}10^{18}$~eV. UHECRs themselves are charged particles, which are subject to a time delay of ${\gtrsim}100-10^5$~years with respect to the time of neutral particles arriving at us, assuming the charged and neutral particles are produced at the same time. However, ultra-high-energy neutrons from Galactic sources~\cite{Anchordoqui:2001nt}, and ultra-high-energy gamma rays from nearby extragalactic sources (within ${\sim}10$ Mpc) may reach the Earth in temporal coincidence with lower-energy EM counterparts, enabling us to identify the \newline UHECR accelerators~\cite{Murase:2009ah,Murase:2011yw,taylor2012}.   

{\it Gravitational wave sources}: The detections of binary black hole (BBH) and binary neutron star (BNS) mergers agree with the predictions of general relativity~\cite{gw170817,firstGW}. The mergers of binary systems are extremely violent events whose main signatures are multi-wavelength EM radiation and GWs, as demonstrated by the observations of GW170817~\cite{gw170817}. Depending on the characteristics of the system, binary neutron star mergers lead to a plethora of post-merger scenarios and counterparts, e.g., short gamma-ray-bursts (SGRBs) or kilonovae, as outlined in~\citep{metzger2012most}. High-energy neutrino production from SGRB jets has been predicted~\cite{Kimura:2017kan,Kimura:2018vvz}. Mergers are expected to be not only strong GW emitters~\cite{firstGW} but also potential sources of cosmic rays, neutrinos, and gamma rays, provided that an accretion disk is present~\cite{Murase:2016etc,Kotera:2016dmp,Moharana:2016xkz}. In addition, various transients, including supernovae and gamma-ray bursts are also promising GW emitters. Searching for neutrino counterparts of GW events can be useful especially for GW sources, in which template or matched-filter analyses are unavailable~\citep[for a review see][]{Bartos:2012vd}.  

{\it Supernovae}:  As demonstrated in the observation of \newline SN1987A~\cite{sn1987a}, core-collapse supernovae (SN) are one of the main candidates for multimessenger transient events. Supernova neutrinos can be detected by current neutrino detectors such as Super-Kamiokande, and SN-induced gravitational waves can be perhaps detected by LIGO, Virgo and KAGRA. Shock breakout emission can be detected from SNe as demonstrated by \textit{Swift} and other telescopes~\cite{swiftSN,Schawinski:2008ba,Gezari:2008jb}. 
Recent observations have shown that that dense circumstellar material is common around the SN progenitor. Thus detection of both MeV and TeV-PeV neutrinos is very promising. MeV neutrinos from nearby supernovae can be detected by various experiments, including IceCube for an event within $D {\leqslant} 50 \,$kpc~\cite{icSN}. 
For Galactic SNe, IceCube can detect 100-1000 or even more high-energy neutrinos, so these objects are effectively multi-energy neutrino sources~\cite{Murase:2017pfe}. 
Furthermore, rare types of SNe may be promising TeV-PeV neutrino emitters. An interesting scenario where only neutrino and GW emissions are possible happens when the SN experiences a choked jet, in which little high-luminosity gamma-ray emission is produced~\cite{Meszaros:2001ms,Senno:2015tsn}. Recent studies have revealed that low-luminosity and ultra-long GRBs are more important as the sources of high-energy neutrinos~\cite{Murase:2006mm,Murase:2013rfa}, in which X-ray counterparts and optical follow-up observations are more critical. 
Super-luminous SNe are often explained by activities of the central engine such as a magnetar. Ions may be accelerated in the magnetar wind, in which neutrinos are naturally produced inside the dense SN ejecta~\cite{Murase:2009pg,Fang:2018hjp}. Fast rotating pulsars are also thought to be interesting objects for GW observations~\citep[e.g.][]{Cutler:2002nw,Stella:2005yz,Kashiyama:2015eua}.   

{\it Long gamma-ray bursts (LGRBs)}: Long gamma-ray bursts are the most luminous explosive phenomena in the universe and thought to be among the most promising sources of UHECRs~\cite{Milgrom:1995um,Waxman:1995vg,Vietri:1995hs}, high-energy neutrinos~\cite{Waxman:1997ti} and GWs~\citep[e.g.][]{Kobayashi:2002by,Suwa:2009si,Ott:2010gv,Kiuchi:2011re}. However, a coincident signal has not yet been found, constraining the contribution of LGRBs to the observed diffuse neutrino flux. 
On the other hand, GeV-TeV neutrino emission is much less constrained and expected to be detectable with IceCube's DeepCore sub-array~\cite{Murase:2013hh,Bartos:2013hf}. At much higher EeV-scale energies, afterglow neutrinos are promising detection candidates for ultra-high-energy neutrino detectors~\cite{Waxman:1999ai,Murase:2006dr}. 
As discussed above, choked jets and low-luminosity GRBs could make a significant contribution to the IceCube flux, and they are interesting targets for neutrino-triggered follow-up observations~\cite{Murase:2006mm}.    

{\it Active galactic nuclei (AGNs)}: These extragalactic objects are galaxies with supermassive black holes in their center which are capable of converting gravitational potential energy and/or rotational energy of the black hole into powerful jets. In the jets, particle acceleration is expected, which means that cosmic rays, gamma rays and neutrinos are produced.  When the jet of the AGN points towards Earth, the AGN is denoted a ``blazar.''  A type of blazar called flat spectrum radio quasars (FSRQs) with prominent external radiation fields are believed to be among the most promising steady sources of PeV-EeV neutrinos~\citep[see][and references therein]{Murase:2015ndr}. 
Blazar flares are also believed to be promising transient neutrino emitters~\citep[e.g.,][]{Dermer:2014vaa}, and coincident events have been reported~\cite{Kadler:2016ygj,nugammaTXS}. 
The power of the multi-messenger observations was successfully demonstrated for the event \mbox{IceCube-170922A}, in which UV, X-ray, and gamma-ray data obtained by follow-up observations (including \textit{Swift} and NuSTAR) were all critical~\cite{Keivani:2018rnh}. This event was transmitted to the community through the AMON network.
Further follow-up or monitoring searches are required to test the physical picture of the blazar TXS 0506+056~\citep[e.g.,][]{Murase:2018iyl,Cerruti:2018tmc,Gao:2018mnu}. 
AGNs without jets can also be promising sources of high-energy neutrinos~\citep[e.g.][]{AlvarezMuniz:2004uz}. Also, the production of secondary particles was predicted for AGNs embedded in galaxy clusters and groups, consistent with the current multi-messenger data~\cite{Fang:2017zjf}.

{\it Other cases: } There are other types of phenomena that can produce multimessenger observations. For example, the disruption of a star by a supermassive black hole, called a tidal disruption event (TDE), can emit neutrinos. UHECRs may be accelerated by TDEs~\cite{Farrar:2008ex,Farrar:2014yla,AlvesBatista:2017shr,Zhang:2017hom}, and high-energy neutrinos may also be produced~\cite{Wang:2015mmh,Dai:2016gtz,Senno:2016bso,Lunardini:2016xwi}. 
GWs can also be emitted especially if the disrupted star is a white dwarf~\cite{Kobayashi:2004py,Haas:2012bk}. 
Other example are soft gamma-ray repeaters (SGRs), which emit large bursts of low energy gamma-rays and hard \mbox{X-rays} at short irregular intervals. They are thought to be isolated neutron stars with very powerful magnetic fields, known as magnetars. Abrupt changes in the magnetic field of the magnetar can accelerate protons or heavy nuclei and produce charged or neutral pions, producing gamma rays or neutrinos~\cite{Ioka:2005er,halzen}. SGRs are also potential sources of GWs~\cite{Ioka:2000hs}.
Fast radio bursts (FRBs) are short radio transients with millisecond durations. The most popular scenarios assume that their progenitors are young neutron stars including magnetars. Magnetic dissipation activity in the nebula may be associated with high-energy gamma-ray flares~\cite{DeLaunay:2016xpf,Murase:2016sqo,Murase:2016ysq}. GW emission is not guaranteed, but can be promising if the FRB progenitors are BNS mergers.
As a final case, AMON will also be able to conduct searches for non-standard physics and new exotic phenomena. For example, primordial black holes (PBHs) are expected to produce a burst of neutrinos, gamma rays, neutrons and protons as they gradually evaporate and then explode in the last seconds of their lives. Sensitivities for PBHs with AMON were presented in~\cite{pbhamon}.
	
%\end{itemize}

%AMON intends to contribute to multimessenger astronomy by looking at multimessenger transient events. 

\begin{table*}
\centering
\footnotesize
\begin{tabular}{|c|c|}
\hline 
Event Class & Expected Messenger Type \\ 
\hline 
\hline 
GRBs & gamma rays, neutrinos, GW, x-rays, IR/O/UV, radio   \\ 
SN & neutrinos, neutrons, x-rays, IR/O/UV   \\ 
Choked SN & neutrinos, GW, x-rays, IR/O/UV, Radio  \\ 
AGNs (Blazars,FSRQs) & gamma rays, neutrinos, x-rays, IR/O/UV, radio   \\ 
PBHs & gamma rays, neutrinos, neutrons  \\
Other (FRBs, TDEs, SGRs) & gamma rays, neutrinos, CRs, GWs, x-rays, IR/O/UV, radio \\

\hline 
\end{tabular} 
\caption{Type of sources and the messengers that can be observed. }

\label{tab:sources}
\end{table*}

%%%%%%%%%%%%%%%
\section{The AMON Framework} \label{sec:frame}

\subsection{Network}
The Astrophysical Multimessenger Observatory Network interconnects astrophysical observatories.  It accepts, stores and analyzes multimessenger data (sub- and above threshold events), and distributes electronic alerts for follow-up observations. 

\begin{figure}[!ht]
\centering
\includegraphics[scale=0.3]{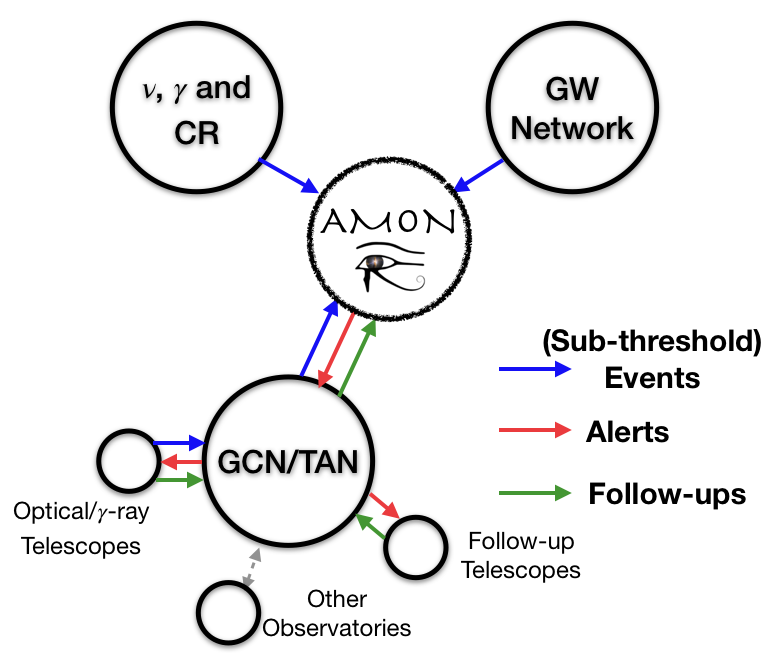}
\caption{The AMON network. Sub-threshold events from different observatories are sent to AMON. Pass-through events and interesting coincident events are sent to GCN for follow-up observations. Interesting follow-up are sent back to AMON (Note that GCN handles many other alert streams independent of AMON).}
\label{fig:network}
\end{figure}

The data shared through the network remains the property of the originating collaborations, and all decisions about data analyses and alert distribution are made by the participating collaborations. For analyses using solely public events, coincidence alerts are made publicly available immediately. Any compelling follow-up alerts, such as observations of new sources or flare events, are sent back to AMON for alert revision and are placed in the AMON archive. The network flow chart is shown in Fig.~\ref{fig:network}.  For more information about data sharing polices see the AMON MoU~\footref{mou}.

The AMON network incorporates {\it triggering} observatories and {\it follow-up} observatories.
\begin{itemize} 
\item Triggering observatories are sensitive to one or more messenger particles and typically offer extensive monitoring capabilities due to their large fields of view and high duty cycles. Examples include the IceCube experiment~\cite{icecube}, the HAWC observatory~\cite{hawc}, the \textit{Fermi}-LAT~\cite{fermi} and \textit{Swift}-BAT satellite telescopes, and the LIGO~\cite{ligo} and VIRGO~\cite{virgo} observatories. 
%They are able to do searches in large portions of the sky and are also able to monitor known sources for transient behavior.
\item Follow-up facilities respond in real-time to AMON alerts and perform observations of the possible multimessenger signal. They include narrow field of view pointed telescopes and telescope networks situated on Earth, such as VERITAS~\cite{veritas} and MASTER~\cite{master}, as well as satellite telescopes like \textit{Swift} XRT/UVOT~\cite{swift}. There might be cases that a follow-up observatory triggers on a relevant event based on, for example, a light curve of a monitored source. These events can also be sent and analyzed through AMON.
\end{itemize}
Partner observatories which have signed the MoU or are considered strong prospects for signing are listed in Table~\ref{tab:partners}. 
\begin{table*}
\centering
\footnotesize
\begin{tabular}{|l|c|c|c|}
\hline 
Observatory & Messenger Types & Role & MoU\\ 
\hline 
\hline 
ANTARES\cite{antares} & neutrinos & triggering  & \checkmark \\ 
FACT\cite{fact} & gamma rays & triggering/follow-up & \checkmark \\ 
\textit{Fermi}-LAT\cite{fermi} & gamma rays & triggering & (public)\\ 
\textit{Fermi}-GBM\cite{fermigbm} & gamma rays & triggering & (public) \\ 
HAWC\cite{hawc} & gamma rays & triggering & \checkmark\\
IceCube\cite{icecube} & neutrinos & triggering & \checkmark \\
MASTER\cite{master} & optical photons & follow-up & \checkmark \\
Auger\cite{auger} & cosmic rays, gamma rays & triggering & \checkmark \\
\textit{Swift}-BAT\cite{swift}& X rays & triggering & (public) \\
\textit{Swift}-XRT\cite{swift} & X rays & follow-up & (public) \\
\textit{Swift}-UVOT\cite{swift} & UV/optical photons & follow-up & (public) \\
VERITAS\cite{veritas} & gamma rays & triggering/follow-up & \checkmark \\
H.E.S.S.\cite{hess} & gamma rays & follow-up & \checkmark \\
MAGIC\cite{magic} & gamma rays & follow-up & \checkmark \\
LMT\cite{LMT} & radio & follow-up & \checkmark \\
LCOGT\cite{LCOGT} & optical & follow-up & \checkmark \\
LIGO\cite{ligo}-Virgo\cite{virgo}* & gravitational waves & triggering & \\
ZTF\cite{ztf} & optical & triggering/follow-up & \\
\hline 
\end{tabular} 
\caption{Triggering and follow-up partner observatories in the AMON Network. \footnotesize{*Ongoing MoU negotiations.}}

\label{tab:partners}
\end{table*}

\subsection{Hardware}
AMON consists of two high-duty cycle servers, with an up-time of 99.9\% percent, corresponding to roughly ten hours of downtime per year. The servers are hosted by the Institute for CyberScience (ICS) at the Pennsylvania State University. The machines are two DELL Model PowerEdge R630 with memory mirroring. These machines are physically and cyber-secure with built-in hardware and power redundancies for high reliability. 

%One of the servers is used as a development machine, were the main AMON software is tested and developed (see section \ref{sec:software}). 

The servers act in concert to manage the real-time system, storing events in the AMON database (see section \ref{sec:database}) and computing the coincidence algorithms.  AMON sends the statistically significant alerts to the Gamma-Ray Coordinates Network/Transient Astronomy Network (GCN/TAN)\footnote{\url{https://gcn.gsfc.nasa.gov/}} by broadcasting the alerts to it. 
AMON also uses a third development machine, a virtual machine located in the Advanced CyberInfrastructure (ACI) system of ICS. This machine is used for the development of code and tests of analyses. 

\subsection{AMON Database}\label{sec:database}

% Add  i/o speed, size, access nodes. 
The AMON system is designed to hold a terabyte-scale data\-base using the 
MySQL\footnote{\url{https://dev.mysql.com/doc/}} as the database management system. The database stores the following information (with database variable names shown in italics): 
\begin{enumerate}[(a)]
\item events received in real-time from the AMON member observatories as well as their archival data: 
\begin{itemize}
    \item \textit{event}: object that holds main information of the event such as position, time, false alarm rate, etc.;
    \item \textit{parameter}: object that holds extra information given by the observatories (e.g. signalness, energy);
    \item \textit{skymap}: object that can hold url of sky maps of the events.
\end{itemize} 
\item AMON alerts: 
\begin{itemize}
    \item \textit{alert}: object that holds the information of a coincidence event that is sent to GCN;
    \item \textit{skymap}: similar as the previous bullet point.
\end{itemize} 
\item observatory configurations: \textit{eventStreamConfig}: detector specifics for a given incoming stream; 
\item cuts used in alert search algorithms: \textit{alertConfig} 
\item multiple analysis stream configurations: 
\begin{itemize}
\item \textit{analysis}: outputs of the analysis needed to rank the alerts;
\item \textit{eventModel}: contextual information for each event contributing to the alert;
\item \textit{source}: type of the astrophysical source;
\item \textit{sourceModel}: information of the astrophysical source spectrum for each messenger type;
\item \textit{messenger}: type of messengers. 
\end{itemize}
\item  \textit{follow-up}: information from follow-up observations of AMON alerts.
\end{enumerate}
%Figure \ref{fig:dbschema} shows the schema of the AMON database.

A performance value that monitors the databases is the buffer cache hit ratio, which measures how often a requested block of data has been queried without requiring disk access. The values of this ratio are 0.998 and 0.983 for the development and production machine, respectively. 

Currently, the database holds IceCube public events from the 40-, 59- and 86-string configurations, the extremely high energy (EHE) and high-energy starting event (HESE) neutrino events, public data from \textit{Fermi}-LAT and \textit{Swift}-BAT, archival data from the Pierre Auger observatory, one year of ANTARES data (from 2008) and current real-time data, HAWC real-time data, and FACT data.

%\begin{figure*}[htp!]
%\centering
%\includegraphics[scale=0.35]{db_schema.png}
%\caption{Relational schema of the AMON database.}
%\label{fig:dbschema}
%\end{figure*}

\subsection{The Software: \texttt{AmonPy}}\label{sec:software}
Following a standard used by many developers in the astrophysics community, the AMON system has been constructed using the Python programming language (version 2.7)\footnote{\url{http://www.python.org}}.  The heart of the AMON software project, \texttt{AmonPy}\footnote{\url{https://github.com/AMONCode/Analysis}; contact the authors to get access.}, has been under development since 2014. 
It relies on three popular Python packages: \texttt{NumPy}~\cite{numpy} for numerical array-based calculations, \texttt{SciPy}~\cite{scipy} for optimization, integration and interpolation algorithms, and \texttt{AstroPy}~\cite{astropy} for time conversions and sky coordinates. It uses \texttt{PyMySQL} to interact with the AMON database, and \texttt{Twisted}\footnote{\url{https://twistedmatrix.com/trac/}}, \texttt{Celery}\footnote{\url{http://www.celeryproject.org/}}, 
and \texttt{RabbitMQ}\footnote{\url{https://www.rabbitmq.com/}} are used for the real-time server application (see Sec.~\ref{sec:rtserver}). 

The software is divided in the following main sub-projects:
\begin{itemize}
\item \texttt{dbase}: contains the main functions that create the data\-base, events, alerts and configurations. It also has the functions to read from the database and write to it.
\item \texttt{analyses}: contains the main algorithms used for the coincidence or pass-through analyses. 
\item \texttt{ops}: contains the definition of the real-time server app and functions that are used for the \texttt{Twisted} plug-ins.
\item \texttt{service}: contains the main \texttt{Twisted} programs that run the real-time server as well as the client functions that can be used by the AMON members to send their events. 
\item \texttt{tools}: contains generic functions, such as distance between events, PSF functions, etc.  that can be used in other projects.
\item \texttt{data}: contains data files needed for the coincidence analyses, such as look-up tables, expected background distributions for each detector or for a coincidence analysis, or any other information given by each observatory that can be useful for an analysis. 
\item \texttt{monitoring}: contains functions that help monitor the performance of the AMON system. 
\end{itemize}

Since new observatories will join AMON, new data streams will be created, and new monitoring tools will be implemented, the software explicitly supports such growth. Naturally, software updates on the production machine do not occur as often as on the development machine. 

\subsubsection{The Real-Time Server}\label{sec:rtserver}
The main application of \texttt{AmonPy} is the real-time server. It handles incoming data, does the coincidence analyses and sends alerts to GCN/TAN. 
The  AMON real-time server is an asynchronous server application based on the \texttt{Twisted}\footnote{\url{https://twistedmatrix.com/trac/}} package, an event-driven network framework developed in Python used to receive data; and the \texttt{Celery}\footnote{\url{http://www.celeryproject.org/}} project, which is able to deal with asynchronous processes by creating several ``workers'' performing different tasks. To communicate with the workers, \texttt{Celery} uses the message-broker \texttt{RabbitMQ}\footnote{\url{https://www.rabbitmq.com/}}.

The server accepts messages in the XML format posted through the HTTPS protocol. The XML format is written in the VOEvent\footnote{\url{https://www.voevent.org}} standard~\cite{voevent}, developed by the International Virtual Observatory Alliance (IVOA). These messages contain the parameters of the sub-threshold data from the member observatories such as observation time, position and its uncertainty, p-value and false alarm rate of the event; the observatories can also add any other information that they consider relevant (e.g., event energy).  The observatories can send their XML files to the AMON servers using a \texttt{Twisted} client application. 
The default parameters that exist in the VOEvent file that are sent to AMON are listed in Table~\ref{tab:voevent}. Extra parameters are easy to incorporate as needed. 
%For extra parameters we add auxiliary parameters as shown in the code snippet below~\ref{fig:auxparam}.

% \lstset{language=XML,
% columns=flexible,
% aboveskip=3mm,
% belowskip=3mm,
% basicstyle=\tiny}
% \begin{minipage}[t]{0.25\linewidth}
% \begin{lstlisting}[caption={Example of an auxiliary parameter in the VOEvent file},captionpos=b]
% <Group name="aux_params">
%             <Param name="energy" dataType="float" value="0.0" ucd="phys.energy" unit="nounits"/>
%             <Param name="signalness" dataType="float" value="0.001" ucd="phys.energy" unit="nounits"/>
% </Group>
% \end{lstlisting}
% \end{minipage}

%\begin{figure*}
%    \centering
%    \includegraphics[scale=0.2]{xml_auxparam.png}
%    \caption{Example of auxiliary parameters in the VOEvent file}
%    \label{fig:auxparam}
%\end{figure*}

\begin{table*}[!htp]
\centering
\footnotesize
\begin{tabular}{|l|c|c|l|}
\hline 
Name & Type &\begin{tabular}{@{}c@{}} VOEvent \\ unified content descriptor\end{tabular} & Description \\            
\hline 
\hline 
stream & int & meta.number & Stream number\\
id & int & meta.number & ID number\\
rev & int & meta.number & Revision number\\
nevents & int & meta.number & Number of events\\
deltaT & float & time.duration & Time window of the search\\
sigmaT & float & time & Uncertainty of the time window\\
false\_pos & float & stat.probability & False positive rate\\
p\_value & float & stat.probability & p-value of the event\\
point\_RA & float & os.eq.ra & Pointing RA \\
point\_Dec & float & os.eq.ra & Pointing Dec \\
psf\_type & & meta.code.multip & Type of PSF \\
\hline 
\end{tabular} 
\caption{Default parameters that appear in the VOEvent file sent by the triggering observatories.}
\label{tab:voevent}
\end{table*}

After receiving the events, the AMON server stores them directly into its database, and then the events are sent to the corresponding workers defined by \texttt{Celery}. 
The workers perform the coincidence analysis between the different data streams. Several workers are configured, depending on the number of analyses planned and agreed upon between the partner observatories and AMON.  After the analysis, any statistically significant coincidences (see Sec.~\ref{sec:channels}) are sent to GCN/TAN using the \textit{Comet}\footnote{\url{https://comet.readthedocs.io/en/stable/}} software package.  Initially those alerts are sent privately between the AMON MoU members. However, if the participant observatories or collaborations agree, the alerts can be made public. Interesting follow-up observations, such as observations of flaring activity or discovery of a new source or sources, are sent back to AMON for alert revision and archival storage.

\subsubsection{The Real-Time Client}\label{sec:rtclient}
Triggering observatories can send their VOEvents with any application using \texttt{Twisted}, or any other software that can provide a HTTPS Post Request with a valid certificate. Inside the AMON project, we have written our own client application using \texttt{Twisted} that can be used by the triggering observatories in case they have not built one yet. The client can be installed during the \texttt{AmonPy} installation process, or for simplicity, the authors also offer an option where the only software needed is the client. 

\subsubsection{Overview of the AMON system with an example}

\begin{figure*}[htp!]
\centering
\includegraphics[scale=0.3]{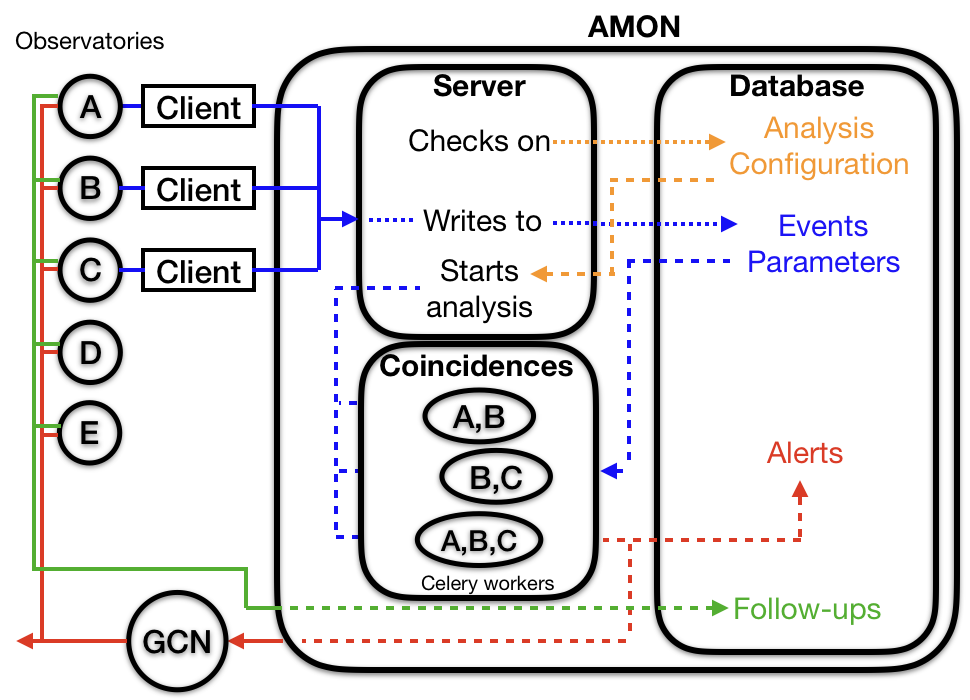}
\caption{The AMON server. Sub-threshold events arrive at the server asynchronously. The server writes the events into the AMON Database and processes the coincidence algorithms. Coincidence results are saved in the database and interesting alerts (e.g., low false alarm rate) are sent to GCN/TAN for either public or private follow-up. Follow-up information is sent back to AMON to keep a record of the observations. }
\label{fig:server}
\end{figure*}

The schematic of the way AMON operates is shown in Fig.~\ref{fig:server}. Triggering observatories (A-C in Fig.~\ref{fig:server}) sent their events to AMON using the real-time client. AMON receives them through the real-time server and stores the events into the database. Event parameters usually have the time and position information of the event, which gets stored in the \textit{event} table. 
Extra parameters  are saved in the \textit{parameters} table. For example, in an analysis involve IceCube data, we will receive energy information which will be stored in this table. 
 The AMON server already knows which coincidence (or pass-through) analyses should be running through the \textit{analysis configuration} table. When a new event has been written in the database, it  is passed to the specific celery worker that does the analysis.%, for example the celery worker that does the B-C analysis in Fig.~\ref{fig:server}. 
Celery workers have access to the database to get the necessary information for the analysis as shown in the figure. The results of the coincidence are saved in the database. If any alert is found that passes a specific criteria (e.g., a false alarm rate of $<1$ per year), it is  sent to AMON members (with D and E acting specifically as follow-ups in Fig.~\ref{fig:server})  for follow-up, as well as the community in general in case the alert was decided to be made public. 

%We would like to do a coincidence analysis between a gamma-ray (A) and a neutrino detector (B). After initial agreements for receiving their data and establishing an analysis, inside AMON we setup the database, so that we have the observatory configurations in place. We give both detectors the \texit{Real-Time Client} so that they can transmit their events to AMON. Once the events start reaching AMON, they are received by the \texit{Real-Time Server}. 
%%%%%%%%%%%%%%%%%%
%%%%%%%%%%%%%%%%%%%%%%%%%%%%%%%%
\section{AMON Analyses and Channels}\label{sec:channels}

\subsection{Coincidence Analyses}

The general technique used for the coincidence analysis will follow a likelihood calculation where we quantify the degree of correlation between different events. An example of this approach is presented in~\cite{colinICLAT}; modifications to it are made depending on the properties of the data and information provided by participating experiments.

A coincidence analysis will usually combine two event streams (see Sec.~\ref{subsec:mmchannels}). From the triggering observatories, AMON receives events that, at minimum, contains the reconstructed position, $\vec{r}_i$, and the time, $t_i$, where $i$ is the event identifier. These variables are not exclusive; AMON can also receive probability density maps of the sky (e.g., from LIGO), or events that are integrated over an extended period of time (e.g., 1-transit events from HAWC).   In addition to the events, each observatory also provides information concerning the distribution of their expected background(s).

If the observatories opt to send additional information on the event quality, it can be added during the likelihood calculation or by using Fisher's method to combine different p-values, producing a $\chi^2$ value~\cite{fisher}. The information could be, for example, the signal-to-noise ratio of the event or the probability of the event to be astrophysical. 

In the case of spatially reconstructed events, we apply spatial (e.g., $\Delta \theta < 3^{\circ}$ between events) and temporal (e.g., a time window of 100~s) cuts to select events. We then construct a (pseudo) log-likelihood statistic. In general the log-likelihood has the form
\begin{equation}
\lambda(\vec{r}_o) = \sum_i \ln{\mathcal{S}(\vec{r}_i,\vec{r}_o)} - \ln{\mathcal{B}(\vec{r}_i)},
\end{equation}
where $\vec{r}_o$ is the joint position of the correlated events,  $\mathcal{S}(\vec{r}_i,\vec{r}_o)$ corresponds to the probability of the signal to come from a region with similar size as the point spread function of the detector, and $\mathcal{B}(\vec{r}_i)$ corresponds to the probability that the event arises from background.  The latter can be a uniform value, or an average over some period of time.  For the temporal component, one can use an open (same probability for all events) or weighted approach (functional weight with respect to the average arrival of the events~\cite{colinICLAT}).

If more information is provided by the observatories an extra term is added to the log-likelihood in the form of $\delta \lambda = \ln{(1-p)/p}$, where $p$ is a value of the quality of the event given by each observatory or we can use Fisher's method.

Generally one maximizes the log-likelihood to find the joint position $\vec{r}_o$. The uncertainty of this position can be determined by combining the uncertainty information of the separate event streams, or can be set to the uncertainty value of the most accurate measurement from all of the participating observatories. 

A key parameter associated with each reported alert is the false alarm rate (FAR) of the coincidence analysis. To obtain the FAR, datasets are scrambled to build a representative distribution of random coincidences. The relevant test statistitc ($\lambda$ or $\chi^2$) value is then used as the statistic to rank coincidences. 
Collaborations contributing to an alert specify the maximum FAR (e.g., $<1$ yr$^{-1}$) at which an alert will be delivered to the GCN/TAN system for further distribution. More restrictive FAR thresholds may also be set by the follow-up observatories.

\subsubsection{Types of Analyses}

\begin{itemize}
\item Archival Analyses: We searched for significant coincident signals in archival data. These studies are crucial, not only to look for signal events in the data, but also to test and improve future tools for real-time coincidence analyses.
Three multimessenger channels have been explored for the archival studies.
In the $\gamma$-$\nu$ channel, we conducted coincidence analyses with the public archival data from IceCube and \textit{Fermi}-LAT observatories. In the study presented in \cite{colinICLAT}, we performed several statistics tests to understand the background and signal distributions. No significant signal excess was observed.

Another example of studies in the same multimessenger channel is the search for gamma-ray flux correlated with high-energy neutrinos from six blazars observed by VERITAS using public IceCube data~\cite{colinBlazar}. No significant excess of neutrinos from the pre-selected blazars was found and 90\% C.L. upper limits were set on the number of expected neutrinos from this search. 

\item Real-time Analyses: At present, the IceCube pass-through for HESE and EHE events are being sent to GCN and are publicly available. A few real-time analysis examples are described in~\cite{icrc2017}.  Also, AMON currently receives events from several observatories. These include IceCube, HAWC, ANTARES, FACT and \textit{Fermi}. Since the coincidence analyses are still under development, the alerts arising from those observatories are not yet available. 

\end{itemize}
% Figure \ref{fig:eventRates} shows the current rate of events for some of the streams. 

% \begin{figure*}[htp!]
% \centering
% \includegraphics[scale=0.45]{logRates.png}
% \caption{Current rates for several streams received by AMON, over a period of 30 days.}
% \label{fig:eventRates}
% \end{figure*}

\subsection{The Multimessenger Channels}\label{subsec:mmchannels}

The coincidence analyses done in AMON are separated into different channels, depending on the type of messenger of each event stream. The events are transmitted to other AMON partners or to the public through GCN/TAN depending on the agreement with the respective team or collaboration. AMON has created so far the $\gamma$ - $\nu$ channel, $\gamma$ - $\nu$ - CR channel, GW - $\gamma$/$\nu$ channel(s) and pass-through channels (see Table~\ref{tab:alerts}). Other characteristics and parameters of the different coincidence streams, including the radius of the best position, latency of the alert and potential sources for discovery are included in the table for completeness. 

%\subsubsection{$\gamma$ - $\nu$ Channel}
%The $\gamma$ - $\nu$ channel looks for coincidences of high-energy gamma rays (or x-rays), and high-energy neutrinos. As mention in the introduction, this channel is mainly focused on the search of the sources of high energy neutrinos. 

%Studies from the IceCube collaboration on archival data showed that, for example, GRBs are not the dominant sources of neutrinos. In the case of the blazar TXS0506+056, it was shown that the combination of different messengers was extremly important in order to identify the blazar as a source of neutrinos. Hence the importance of this AMON channel. 
%The current coincidence streams that are under development are: HAWC/IceCube, Swift-BAT/IceCube, \textit{Fermi}-LAT/IceCube, and \textit{Fermi}-LAT/ANTARES. 

\begin{table*}
\centering
\footnotesize
\begin{tabular}{ |l|l|l|l|l|l|}%l| }
\hline 
Channel & Facilities & $\delta r$ & $\Delta T_{search}$  & \begin{tabular}{@{}c@{}} Latency \\ (hours)\end{tabular} &\begin{tabular}{@{}c@{}}Potential \\ Sources\end{tabular}\\% & Status  \\            
\hline 
\hline
\multirow{4}{*}{$\gamma$-$\nu$}& ANTARES-{\it Fermi}-LAT &$\sim$0.3$^{\circ}$ & 2000 sec & 1-12 & GRBs \\%& \checkmark \\ 
%\hline 
& IceCube-HAWC & $\sim$0.1$^{\circ}$ & $\sim$6 hours & 3-8  & AGNs, GRBs \\%& \checkmark \\ 
%\hline 
& IceCube-{\it Fermi}-LAT & $\sim$0.3$^{\circ}$ & 2000 sec & 1-12 & GRBs \\%& \checkmark \\ 
%\hline 
& IceCube-\textit{Swift}-BAT & $<$0.1$^{\circ}$ & 300 sec &1-8 & \begin{tabular}{@{}c@{}} AGNs, GRBs \\ TDEs, SGRs\end{tabular} \\%& \\ 
\hline 
\multirow{3}{*}{$\gamma$-GW}& LIGO/Virgo- HAWC & $\leq$0.8$^{\circ}$ & $\sim$6 hours & 3-8  & GRBs \\%& \\ 
%\hline 
& LIGO/Virgo-{\it Fermi}-LAT & $\sim$0.3$^{\circ}$ & 2000 sec & 1-12 & GRBs \\%& \\ 
%\hline 
& LIGO/Virgo-\textit{Swift}-BAT & $<$0.1$^{\circ}$ & 300 sec &1-8 & \begin{tabular}{@{}c@{}} GRBs \\ TDEs, SGRs\end{tabular} \\%& \\ 
\hline 
$\gamma$-$\nu$-CR & IceCube-HAWC-Pierre Auger & $\leq$0.8$^{\circ}$ & 2000 sec & 1-12 & PBHs \\%& \\ 
\hline 
\multirow{3}{*}{Pass-through}& HESE-EHE IceCube & $<$0.75$^{\circ}$ (90\%) & -- & $<1$ min  & AGNs, GRBs \\%& Ended\\ 
%\hline
& Gold-Bronze IceCube & $<$0.4$^{\circ}$ (90\%) & -- & $<1$ min  & AGNs, GRBs \\%& \checkmark \\
%\hline 
& HAWC Burst & $\leq$0.8$^{\circ}$ (68\%) & 0.2,1,10,100 sec & $<1$ min & GRBs \\%& \checkmark \\ 
%\hline 
& FACT & $<$0.1$^{\circ}$ & -- & $<$1 min & \begin{tabular}{@{}c@{}} AGNs, GRBs \\ TDEs, SGRs\end{tabular} \\%& \\
& Auger Doublets & $\sim$1$^{\circ}$ & -- & $\lesssim$10min & \begin{tabular}{@{}c@{}} AGNs, GRBs \\ TDEs, SGRs\end{tabular} \\%& \\
\hline 
\end{tabular} 
\caption{Characteristics of different AMON alerts that are in development. The table includes the angular uncertainty of the best position, the latency of the alert and possible sources that can be monitored. For coincident events, the angular uncertainty is dominated by the instrument with better angular resolution. For pass-through streams, the containment radius is shown in parentheses. Latency is the estimated time taken by the observatories to process the events. IceCube-HAWC and IceCube-{\it Fermi} LAT are under collaobration review. The HESE and EHE streams will end in 2019 and be substituted by the IceCube ``Gold'' and ``Bronze'' streams. HAWC Burst alerts will start being sent in 2019.}
\label{tab:alerts}
\end{table*}

%\subsubsection{$\gamma$ - $\nu$ - Cosmic-Ray Channel}
%An example of an AMON-brokered multimessenger coincidence search is the analysis of neutrino data from IceCube, gamma-ray data from HAWC and cosmic ray data from the Pierre Auger Observatory. The aim of this search is for evaporation signature of a primordial black hole (PBH) \cite{pbhs,pbhamon}. 
%\subsubsection{GW channels}
%The science case for searches of transients emitting both gravitational waves and EM radiation, and also correlated high-energy neutrinos was validated after the GW170817/GRB170817a event \cite{gw170817}. With the advancement of LIGO and Virgo, coincidence searches between those three messenger is the logical step to take. 
%There are on-going negotiations between AMON and the LIGO-Virgo Scientific Collaboration to obtain sub-threshold data from previous observation runs O1\& O2, as well as for the observation runs to come.

%We are developing archival and real-time coincidence analysis between EM observations ranging from X-ray to very-high energies done by Swift-BAT/Fermi-GBM/HAWC and gravitational waves detected from LIGO-Virgo.

\subsection{Pass-through Channel}
There are cases in which triggering observatories wish to distribute individual events that are near or above threshold, which allows the observatory to make a stand-alone claim for discovery. In these cases, AMON can serve as a conduit for propagating the pertinent event information as an alert, enabling other observatories to do timely follow-up observations. 
This has been implemented for two IceCube streams, the HESE and EHE streams, operating since April and July of 2016, respectively. Optical, X-ray and gamma-ray telescopes have been following these two streams. This channel led to a series of follow-ups including the ones presented by the AMON team at the ICRC 2017~\cite{icrc2017}, where the \textit{Swift} telescope followed-up three HESE and one EHE events.  A highly important event in the pass-through category is the IceCube event \mbox{IceCube-170922A}, which gave us the first hint of a neutrino source from a blazar.\footnote{The AMON GCN alerts can be found in \url{https://gcn.gsfc.nasa.gov/gcn/amon_hese_events.html} and \url{https://gcn.gsfc.nasa.gov/gcn/amon_ehe_events.html}}

New pass-through channels are in the planning stage. This include a new IceCube stream that focus on sending alerts to the gamma-ray community, and a new HAWC stream, based on a search for GRB-like events, which can also be used to send alerts to the high-energy astrophysics community. In the near future, AMON also plans to include FACT alerts and Auger doublet-event alerts.

%%%%%%%%%%%%%%%%%%
\section{Current Status}\label{sec:results}
%In this section we present the current status of the AMON Project. 
At present, the accomplishments of AMON include:
\begin{itemize}
    \item Fast distribution of IceCube alerts of likely cosmic neutrinos to GCN/TAN.
    \item Helping catalyze multiple follow-up campaigns for likely-cosmic neutrinos including \mbox{IceCube-170922A} event~\cite{nugammaTXS}.
    \item Performing several archival searches for transient and flaring $\gamma$-$\nu$ and $\nu$-CR multimessenger sources.
\end{itemize}
AMON has been distributing IceCube alerts via GCN since April of 2016. 
We expect to start issuing $\gamma-\nu$ coincidence alerts using the combined HAWC+IceCube and \textit{Fermi}-LAT+ANTARES detector combinations, as well as implementing pass-through for HAWC alerts to GCN/TAN. 
Currently, the total number of events that AMON receives from different observatories is on the order of $\sim10^3$ per day. 
%We will review these examples in the following sections.

%%%%%%%%%%%%%%%%%%
\section{Summary}\label{sec:concl}
The field of multimessenger astroparticle physics is growing rapidly and benefits strongly from the cooperation of observatories. The AMON cyber-infrastructure is explicitly tailored for this important task, enabling real-time and archival searches for multimessenger sources using sub-threshold events from multiple observatories. AMON will distribute alerts ---publicly or privately--- with minimal latency to follow-up recipients, enabling identification of transient astrophysical phenomena, providing the multimessenger astronomy community with new pathways to discovery.

%%%%%%%%%%%%%%%%%%
%\appendix
%\section{AMON Client Code}

\section*{Acknowledgments}
\small{
This research or portions of this research were conducted with Advanced CyberInfrastructure computational resources provided by the Institute for CyberScience at the Pennsylvania State University (\url{http://ics.psu.edu}).

This material is based upon work supported by the National Science Foundation under Grants PHY-1708146 and PHY-1806854 and by the Institute for Gravitation and the Cosmos of the Pennsylvania State University. Any opinions, findings, and conclusions or recommendations expressed in this material are those of the author(s) and do not necessarily reflect the views of the National Science Foundation.}

\section*{References}

\interlinepenalty=10000
\bibliography{biblio}

\begin{thebibliography}{100}
\expandafter\ifx\csname url\endcsname\relax
  \def\url#1{\texttt{#1}}\fi
\expandafter\ifx\csname urlprefix\endcsname\relax\def\urlprefix{URL }\fi
\expandafter\ifx\csname href\endcsname\relax
  \def\href#1#2{#2} \def\path#1{#1}\fi

\bibitem{sn1987a}
K.~Hirata, T.~Kajita, M.~Koshiba, et~al., {Observation of a Neutrino Burst from
  the Supernova SN1987A}, Phys. Rev. Lett. 58 (1987) 1490--1494.

\bibitem{ic2017}
I.~Collaboration, ASAS-SN, A.~Team, et~al., {Multiwavelength follow-up of a
  rare IceCube neutrino multiplet}, A\&A 607 (2017) A115.

\bibitem{icrc2017}
A.~Keivani, D.~F. Cowen, D.~B. Fox, et~al., {Four Swift searches for Transient
  Sources of High-Energy Neutrinos}, PoS ICRC 2017 (2017) 8.

\bibitem{nugammaTXS}
T.~I. Collaboration, Fermi-LAT, MAGIC, et~al., {Multimessenger observations of
  a flaring blazar coincident with high-energy neutrino IceCube-170922A},
  Science 361 (2018) 8.

\bibitem{ICFlare}
{IceCube Collaboration}, Neutrino emission from the direction of the blazar txs
  0506+056 prior to the icecube-170922a alert, Science 361 (2018) 147--151.

\bibitem{Keivani:2018rnh}
A.~Keivani, et~al., {A Multimessenger Picture of the Flaring Blazar TXS
  0506+056: implications for High-Energy Neutrino Emission and Cosmic Ray
  Acceleration}, Astrophys. J. 864~(1) (2018) 84.
\newblock \href {http://arxiv.org/abs/1807.04537} {\path{arXiv:1807.04537}},
  \href {http://dx.doi.org/10.3847/1538-4357/aad59a}
  {\path{doi:10.3847/1538-4357/aad59a}}.

\bibitem{firstGW}
B.~Abbott, R.~Abbott, T.~Abbot, et~al., {Observation of Gravitational Waves
  from a Binary Black Hole Merger}, Phys. Rev. Lett. 116 (2016) 061102.

\bibitem{gw170817}
L.~S. Collaboration, V.~Collaboration, et~al., {Multi-messenger Observations of
  a Binary Neutron Star Merger}, ApJL 848 (2017) 59.

\bibitem{amon2013}
M.~Smith, et~al., {The Astrophysical Multimessenger Observatory Network
  (AMON)}, Astroparticle Physics 45 (2013) 56--70.

\bibitem{amonicrc}
A.~{Keivani}, H.~{Ayala}, J.~{DeLaunay}, {AMON Core Team}, {Astrophysical
  Multimessenger Observatory Network (AMON): Science, Infrastructure, and
  Status}, PoS ICRC 2017 35 (2017) 629.
\newblock \href {http://arxiv.org/abs/1708.04724} {\path{arXiv:1708.04724}}.

\bibitem{gcn}
S.~Barthelmy, et~al., {BACODINE, the Real-Time BATSE Gamma-Ray Burst
  Coordinates Distribution Network}, Astrophysics and Space Science 231 (1995)
  235--238.

\bibitem{originCR}
J.~Linsley, {Evidence for a primary cosmic-ray particle with energy $10^{20}$
  eV}, Phys. Rev. Lett. 10 (1963) 146--148.

\bibitem{icnature}
F.~Halzen, High-energy neutrino astrophysics, Nat. Phys 13 (2016) 232--238.

\bibitem{firstpevnu}
M.~Aarsten, et~al., {First Observation of PeV-energy neutrions with IceCube},
  Phys. Rev. Lett. 111 (2013) 021103.

\bibitem{Aartsen:2013jdh}
M.~G. Aartsen, et~al., {Evidence for High-Energy Extraterrestrial Neutrinos at
  the IceCube Detector}, Science 342 (2013) 1242856.
\newblock \href {http://arxiv.org/abs/1311.5238} {\path{arXiv:1311.5238}},
  \href {http://dx.doi.org/10.1126/science.1242856}
  {\path{doi:10.1126/science.1242856}}.

\bibitem{cosmicnu}
M.~Aarsten, et~al., {Observation and characterization of a cosmic muon neutrino
  flux from the northern hemisphere using six years of IceCube data}, ApJ 833
  (2016) 18.

\bibitem{Aartsen:2014muf}
M.~G. Aartsen, et~al., {Atmospheric and astrophysical neutrinos above 1 TeV
  interacting in IceCube}, Phys. Rev. D91~(2) (2015) 022001.
\newblock \href {http://arxiv.org/abs/1410.1749} {\path{arXiv:1410.1749}},
  \href {http://dx.doi.org/10.1103/PhysRevD.91.022001}
  {\path{doi:10.1103/PhysRevD.91.022001}}.

\bibitem{Aartsen:2015knd}
M.~G. Aartsen, et~al., {A combined maximum-likelihood analysis of the
  high-energy astrophysical neutrino flux measured with IceCube}, Astrophys. J.
  809~(1) (2015) 98.
\newblock \href {http://arxiv.org/abs/1507.03991} {\path{arXiv:1507.03991}},
  \href {http://dx.doi.org/10.1088/0004-637X/809/1/98}
  {\path{doi:10.1088/0004-637X/809/1/98}}.

\bibitem{Murase:2015xka}
K.~Murase, D.~Guetta, M.~Ahlers, {Hidden Cosmic-Ray Accelerators as an Origin
  of TeV-PeV Cosmic Neutrinos}, Phys. Rev. Lett. 116~(7) (2016) 071101.
\newblock \href {http://arxiv.org/abs/1509.00805} {\path{arXiv:1509.00805}},
  \href {http://dx.doi.org/10.1103/PhysRevLett.116.071101}
  {\path{doi:10.1103/PhysRevLett.116.071101}}.

\bibitem{kmwaxman}
K.~Murase, E.~Waxman, {Constraining high-energy cosmic neutrino sources:
  Implications and prospects}, Phys. Rev. D 94 (2016) 103006.

\bibitem{hillas}
A.~Hillas, {The origin of ultra-high-energy cosmic rays}, Ann. Rev. Astron.
  Astrophys. 22 (1984) 425--444.

\bibitem{Fang:2017zjf}
K.~Fang, K.~Murase, {Linking High-Energy Cosmic Particles by Black Hole Jets
  Embedded in Large-Scale Structures}, Nature Phys. 14 (2018) 396, [Nature
  Phys.14,no.4,396(2018)].
\newblock \href {http://arxiv.org/abs/1704.00015} {\path{arXiv:1704.00015}},
  \href {http://dx.doi.org/10.1038/s41567-017-0025-4}
  {\path{doi:10.1038/s41567-017-0025-4}}.

\bibitem{Anchordoqui:2001nt}
L.~A. Anchordoqui, H.~Goldberg, T.~J. Weiler, {An Auger test of the Cen A model
  of highest energy cosmic rays}, Phys. Rev. Lett. 87 (2001) 081101.
\newblock \href {http://arxiv.org/abs/astro-ph/0103043}
  {\path{arXiv:astro-ph/0103043}}, \href
  {http://dx.doi.org/10.1103/PhysRevLett.87.081101}
  {\path{doi:10.1103/PhysRevLett.87.081101}}.

\bibitem{Murase:2009ah}
K.~Murase, {Ultrahigh-Energy Photons as a Probe of Nearby Transient
  Ultrahigh-Energy Cosmic-Ray Sources and Possible Lorentz-Invariance
  Violation}, Phys. Rev. Lett. 103 (2009) 081102.
\newblock \href {http://arxiv.org/abs/0904.2087} {\path{arXiv:0904.2087}},
  \href {http://dx.doi.org/10.1103/PhysRevLett.103.081102}
  {\path{doi:10.1103/PhysRevLett.103.081102}}.

\bibitem{Murase:2011yw}
K.~Murase, {High-Energy Emission Induced by Ultra-high-Energy Photons as a
  Probe of Ultra-high-Energy Cosmic-Ray Accelerators Embedded in the Cosmic
  Web}, Astrophys. J. 745 (2012) L16.
\newblock \href {http://arxiv.org/abs/1111.0936} {\path{arXiv:1111.0936}},
  \href {http://dx.doi.org/10.1088/2041-8205/745/2/L16}
  {\path{doi:10.1088/2041-8205/745/2/L16}}.

\bibitem{taylor2012}
A.~M. {Taylor}, J.~A. {Hinton}, P.~{Blasi}, M.~{Ave}, {Identifying Nearby
  Accelerators of Ultrahigh Energy Cosmic Rays Using Ultrahigh Energy (and Very
  High Energy) Photons}, Phys. Rev. Lett. 103 (2009) 051102.

\bibitem{metzger2012most}
B.~D. Metzger, E.~Berger, What is the most promising electromagnetic
  counterpart of a neutron star binary merger?, The Astrophysical Journal
  746~(1) (2012) 48.

\bibitem{Kimura:2017kan}
S.~S. Kimura, K.~Murase, P.~Mészáros, K.~Kiuchi, {High-Energy Neutrino
  Emission from Short Gamma-Ray Bursts: Prospects for Coincident Detection with
  Gravitational Waves}, Astrophys. J. 848~(1) (2017) L4.
\newblock \href {http://arxiv.org/abs/1708.07075} {\path{arXiv:1708.07075}},
  \href {http://dx.doi.org/10.3847/2041-8213/aa8d14}
  {\path{doi:10.3847/2041-8213/aa8d14}}.

\bibitem{Kimura:2018vvz}
S.~S. Kimura, K.~Murase, I.~Bartos, K.~Ioka, I.~S. Heng, P.~Mészáros,
  {Transejecta high-energy neutrino emission from binary neutron star mergers},
  Phys. Rev. D98~(4) (2018) 043020.
\newblock \href {http://arxiv.org/abs/1805.11613} {\path{arXiv:1805.11613}},
  \href {http://dx.doi.org/10.1103/PhysRevD.98.043020}
  {\path{doi:10.1103/PhysRevD.98.043020}}.

\bibitem{Murase:2016etc}
K.~Murase, K.~Kashiyama, P.~Mészáros, I.~Shoemaker, N.~Senno, {Ultrafast
  Outflows from Black Hole Mergers with a Minidisk}, Astrophys. J. 822~(1)
  (2016) L9.
\newblock \href {http://arxiv.org/abs/1602.06938} {\path{arXiv:1602.06938}},
  \href {http://dx.doi.org/10.3847/2041-8205/822/1/L9}
  {\path{doi:10.3847/2041-8205/822/1/L9}}.

\bibitem{Kotera:2016dmp}
K.~Kotera, J.~Silk, {Ultrahigh Energy Cosmic Rays and Black Hole Mergers},
  Astrophys. J. 823~(2) (2016) L29.
\newblock \href {http://arxiv.org/abs/1602.06961} {\path{arXiv:1602.06961}},
  \href {http://dx.doi.org/10.3847/2041-8205/823/2/L29}
  {\path{doi:10.3847/2041-8205/823/2/L29}}.

\bibitem{Moharana:2016xkz}
R.~Moharana, S.~Razzaque, N.~Gupta, P.~Meszaros, {High Energy Neutrinos from
  the Gravitational Wave event GW150914 possibly associated with a short
  Gamma-Ray Burst}, Phys. Rev. D93~(12) (2016) 123011.
\newblock \href {http://arxiv.org/abs/1602.08436} {\path{arXiv:1602.08436}},
  \href {http://dx.doi.org/10.1103/PhysRevD.93.123011}
  {\path{doi:10.1103/PhysRevD.93.123011}}.

\bibitem{Bartos:2012vd}
I.~Bartos, P.~Brady, S.~Marka, {How Gravitational-wave Observations Can Shape
  the Gamma-ray Burst Paradigm}, Class. Quant. Grav. 30 (2013) 123001.
\newblock \href {http://arxiv.org/abs/1212.2289} {\path{arXiv:1212.2289}},
  \href {http://dx.doi.org/10.1088/0264-9381/30/12/123001}
  {\path{doi:10.1088/0264-9381/30/12/123001}}.

\bibitem{swiftSN}
A.~M. Soderberg, et~al., {An Extremely Luminous X-ray Outburst Marking the
  Birth of a Normal Supernova}, Nature 453 (2008) 469--474.
\newblock \href {http://arxiv.org/abs/0802.1712} {\path{arXiv:0802.1712}},
  \href {http://dx.doi.org/10.1038/nature06997}
  {\path{doi:10.1038/nature06997}}.

\bibitem{Schawinski:2008ba}
K.~Schawinski, et~al., {Supernova Shock Breakout from a Red Supergiant},
  Science 321 (2008) 223.
\newblock \href {http://arxiv.org/abs/0803.3596} {\path{arXiv:0803.3596}},
  \href {http://dx.doi.org/10.1126/science.1160456}
  {\path{doi:10.1126/science.1160456}}.

\bibitem{Gezari:2008jb}
S.~Gezari, et~al., {Probing Shock Breakout with Serendipitous GALEX Detections
  of Two SNLS Type II-P Supernovae}, Astrophys. J. 683 (2008) L131.
\newblock \href {http://arxiv.org/abs/0804.1123} {\path{arXiv:0804.1123}},
  \href {http://dx.doi.org/10.1086/591647} {\path{doi:10.1086/591647}}.

\bibitem{icSN}
R.~Abbasi, Y.~Abdou, T.~Abu-Zayyad, et~al., {IceCube sensitivity for low-energy
  neutrinos from nearby supernovae}, Astron. Astrophys. 535 (2011) 18.

\bibitem{Murase:2017pfe}
K.~Murase, {New Prospects for Detecting High-Energy Neutrinos from Nearby
  Supernovae}, Phys. Rev. D97~(8) (2018) 081301.
\newblock \href {http://arxiv.org/abs/1705.04750} {\path{arXiv:1705.04750}},
  \href {http://dx.doi.org/10.1103/PhysRevD.97.081301}
  {\path{doi:10.1103/PhysRevD.97.081301}}.

\bibitem{Meszaros:2001ms}
P.~Meszaros, E.~Waxman, {TeV neutrinos from successful and choked gamma-ray
  bursts}, Phys. Rev. Lett. 87 (2001) 171102.
\newblock \href {http://arxiv.org/abs/astro-ph/0103275}
  {\path{arXiv:astro-ph/0103275}}, \href
  {http://dx.doi.org/10.1103/PhysRevLett.87.171102}
  {\path{doi:10.1103/PhysRevLett.87.171102}}.

\bibitem{Senno:2015tsn}
N.~Senno, K.~Murase, P.~Meszaros, {Choked Jets and Low-Luminosity Gamma-Ray
  Bursts as Hidden Neutrino Sources}, Phys. Rev. D93~(8) (2016) 083003.
\newblock \href {http://arxiv.org/abs/1512.08513} {\path{arXiv:1512.08513}},
  \href {http://dx.doi.org/10.1103/PhysRevD.93.083003}
  {\path{doi:10.1103/PhysRevD.93.083003}}.

\bibitem{Murase:2006mm}
K.~Murase, K.~Ioka, S.~Nagataki, T.~Nakamura, {High Energy Neutrinos and
  Cosmic-Rays from Low-Luminosity Gamma-Ray Bursts?}, Astrophys. J. 651 (2006)
  L5--L8.
\newblock \href {http://arxiv.org/abs/astro-ph/0607104}
  {\path{arXiv:astro-ph/0607104}}, \href {http://dx.doi.org/10.1086/509323}
  {\path{doi:10.1086/509323}}.

\bibitem{Murase:2013rfa}
K.~Murase, M.~Ahlers, B.~C. Lacki, {Testing the Hadronuclear Origin of PeV
  Neutrinos Observed with IceCube}, Phys. Rev. D88~(12) (2013) 121301.
\newblock \href {http://arxiv.org/abs/1306.3417} {\path{arXiv:1306.3417}},
  \href {http://dx.doi.org/10.1103/PhysRevD.88.121301}
  {\path{doi:10.1103/PhysRevD.88.121301}}.

\bibitem{Murase:2009pg}
K.~Murase, P.~Meszaros, B.~Zhang, {Probing the birth of fast rotating magnetars
  through high-energy neutrinos}, Phys. Rev. D79 (2009) 103001.
\newblock \href {http://arxiv.org/abs/0904.2509} {\path{arXiv:0904.2509}},
  \href {http://dx.doi.org/10.1103/PhysRevD.79.103001}
  {\path{doi:10.1103/PhysRevD.79.103001}}.

\bibitem{Fang:2018hjp}
K.~Fang, B.~D. Metzger, K.~Murase, I.~Bartos, K.~Kotera, {Multimessenger
  Implications of AT2018cow: High-Energy Cosmic Ray and Neutrino Emissions from
  Magnetar-Powered Super-Luminous Transients}, Astrophys. J. 878 (2019) 34.
\newblock \href {http://arxiv.org/abs/1812.11673} {\path{arXiv:1812.11673}}.

\bibitem{Cutler:2002nw}
C.~Cutler, {Gravitational waves from neutron stars with large toroidal B
  fields}, Phys. Rev. D66 (2002) 084025.
\newblock \href {http://arxiv.org/abs/gr-qc/0206051}
  {\path{arXiv:gr-qc/0206051}}, \href
  {http://dx.doi.org/10.1103/PhysRevD.66.084025}
  {\path{doi:10.1103/PhysRevD.66.084025}}.

\bibitem{Stella:2005yz}
L.~Stella, S.~Dall'Osso, G.~Israel, A.~Vecchio, {Gravitational radiation from
  newborn magnetars}, Astrophys. J. 634 (2005) L165--L168.
\newblock \href {http://arxiv.org/abs/astro-ph/0511068}
  {\path{arXiv:astro-ph/0511068}}, \href {http://dx.doi.org/10.1086/498685}
  {\path{doi:10.1086/498685}}.

\bibitem{Kashiyama:2015eua}
K.~Kashiyama, K.~Murase, I.~Bartos, K.~Kiuchi, R.~Margutti, {Multi-Messenger
  Tests for Fast-Spinning Newborn Pulsars Embedded in Stripped-Envelope
  Supernovae}, Astrophys. J. 818~(1) (2016) 94.
\newblock \href {http://arxiv.org/abs/1508.04393} {\path{arXiv:1508.04393}},
  \href {http://dx.doi.org/10.3847/0004-637X/818/1/94}
  {\path{doi:10.3847/0004-637X/818/1/94}}.

\bibitem{Milgrom:1995um}
M.~Milgrom, V.~Usov, {Possible association of ultrahigh-energy cosmic ray
  events with strong gamma-ray bursts}, Astrophys. J. 449 (1995) L37.
\newblock \href {http://arxiv.org/abs/astro-ph/9505009}
  {\path{arXiv:astro-ph/9505009}}, \href {http://dx.doi.org/10.1086/309633}
  {\path{doi:10.1086/309633}}.

\bibitem{Waxman:1995vg}
E.~Waxman, {Cosmological gamma-ray bursts and the highest energy cosmic rays},
  Phys. Rev. Lett. 75 (1995) 386--389.
\newblock \href {http://arxiv.org/abs/astro-ph/9505082}
  {\path{arXiv:astro-ph/9505082}}, \href
  {http://dx.doi.org/10.1103/PhysRevLett.75.386}
  {\path{doi:10.1103/PhysRevLett.75.386}}.

\bibitem{Vietri:1995hs}
M.~Vietri, {On the acceleration of ultrahigh-energy cosmic rays in gamma-ray
  bursts}, Astrophys. J. 453 (1995) 883--889.
\newblock \href {http://arxiv.org/abs/astro-ph/9506081}
  {\path{arXiv:astro-ph/9506081}}, \href {http://dx.doi.org/10.1086/176448}
  {\path{doi:10.1086/176448}}.

\bibitem{Waxman:1997ti}
E.~Waxman, J.~N. Bahcall, {High-energy neutrinos from cosmological gamma-ray
  burst fireballs}, Phys. Rev. Lett. 78 (1997) 2292--2295.
\newblock \href {http://arxiv.org/abs/astro-ph/9701231}
  {\path{arXiv:astro-ph/9701231}}, \href
  {http://dx.doi.org/10.1103/PhysRevLett.78.2292}
  {\path{doi:10.1103/PhysRevLett.78.2292}}.

\bibitem{Kobayashi:2002by}
S.~Kobayashi, P.~Meszaros, {Gravitational radiation from gamma-ray burst
  progenitors}, Astrophys. J. 589 (2003) 861--870.
\newblock \href {http://arxiv.org/abs/astro-ph/0210211}
  {\path{arXiv:astro-ph/0210211}}, \href {http://dx.doi.org/10.1086/374733}
  {\path{doi:10.1086/374733}}.

\bibitem{Suwa:2009si}
Y.~Suwa, K.~Murase, {Probing the central engine of long gamma-ray bursts and
  hypernovae with gravitational waves}, Phys. Rev. D80 (2009) 123008.
\newblock \href {http://arxiv.org/abs/0906.3833} {\path{arXiv:0906.3833}},
  \href {http://dx.doi.org/10.1103/PhysRevD.80.123008}
  {\path{doi:10.1103/PhysRevD.80.123008}}.

\bibitem{Ott:2010gv}
C.~D. Ott, C.~Reisswig, E.~Schnetter, E.~O'Connor, U.~Sperhake, F.~Loffler,
  P.~Diener, E.~Abdikamalov, I.~Hawke, A.~Burrows, {Dynamics and Gravitational
  Wave Signature of Collapsar Formation}, Phys. Rev. Lett. 106 (2011) 161103.
\newblock \href {http://arxiv.org/abs/1012.1853} {\path{arXiv:1012.1853}},
  \href {http://dx.doi.org/10.1103/PhysRevLett.106.161103}
  {\path{doi:10.1103/PhysRevLett.106.161103}}.

\bibitem{Kiuchi:2011re}
K.~Kiuchi, M.~Shibata, P.~J. Montero, J.~A. Font, {Gravitational waves from the
  Papaloizou-Pringle instability in black hole-torus systems}, Phys. Rev. Lett.
  106 (2011) 251102.
\newblock \href {http://arxiv.org/abs/1105.5035} {\path{arXiv:1105.5035}},
  \href {http://dx.doi.org/10.1103/PhysRevLett.106.251102}
  {\path{doi:10.1103/PhysRevLett.106.251102}}.

\bibitem{Murase:2013hh}
K.~Murase, K.~Kashiyama, P.~Mészáros, {Subphotospheric Neutrinos from
  Gamma-Ray Bursts: The Role of Neutrons}, Phys. Rev. Lett. 111 (2013) 131102.
\newblock \href {http://arxiv.org/abs/1301.4236} {\path{arXiv:1301.4236}},
  \href {http://dx.doi.org/10.1103/PhysRevLett.111.131102}
  {\path{doi:10.1103/PhysRevLett.111.131102}}.

\bibitem{Bartos:2013hf}
I.~Bartos, A.~M. Beloborodov, K.~Hurley, S.~Márka, {Detection Prospects for
  GeV Neutrinos from Collisionally Heated Gamma-ray Bursts with
  IceCube/DeepCore}, Phys. Rev. Lett. 110~(24) (2013) 241101.
\newblock \href {http://arxiv.org/abs/1301.4232} {\path{arXiv:1301.4232}},
  \href {http://dx.doi.org/10.1103/PhysRevLett.110.241101}
  {\path{doi:10.1103/PhysRevLett.110.241101}}.

\bibitem{Waxman:1999ai}
E.~Waxman, J.~N. Bahcall, {Neutrino afterglow from gamma-ray bursts: Similar to
  10**18-eV}, Astrophys. J. 541 (2000) 707--711.
\newblock \href {http://arxiv.org/abs/hep-ph/9909286}
  {\path{arXiv:hep-ph/9909286}}, \href {http://dx.doi.org/10.1086/309462}
  {\path{doi:10.1086/309462}}.

\bibitem{Murase:2006dr}
K.~Murase, S.~Nagataki, {High Energy Neutrino Flash from Far-UV/X-ray Flares of
  Gamma-Ray Bursts}, Phys. Rev. Lett. 97 (2006) 051101.
\newblock \href {http://arxiv.org/abs/astro-ph/0604437}
  {\path{arXiv:astro-ph/0604437}}, \href
  {http://dx.doi.org/10.1103/PhysRevLett.97.051101}
  {\path{doi:10.1103/PhysRevLett.97.051101}}.

\bibitem{Murase:2015ndr}
K.~{Murase}, {Active Galactic Nuclei as High-Energy Neutrino Sources}, in:
  Neutrino Astronomy: Current Status, Future Prospects. Edited by Thomas
  Gaisser Albrecht Karle. Published by World Scientific Publishing Co. Pte.
  Ltd., 2017. ISBN \#9789814759410, pp. 15-31, World Scientific Publishing Co.,
  2017, pp. 15--31.
\newblock \href {http://dx.doi.org/10.1142/9789814759410_0002}
  {\path{doi:10.1142/9789814759410_0002}}.

\bibitem{Dermer:2014vaa}
C.~D. Dermer, K.~Murase, Y.~Inoue, {Photopion Production in Black-Hole Jets and
  Flat-Spectrum Radio Quasars as PeV Neutrino Sources}, JHEAp 3-4 (2014)
  29--40.
\newblock \href {http://arxiv.org/abs/1406.2633} {\path{arXiv:1406.2633}},
  \href {http://dx.doi.org/10.1016/j.jheap.2014.09.001}
  {\path{doi:10.1016/j.jheap.2014.09.001}}.

\bibitem{Kadler:2016ygj}
M.~Kadler, et~al., {Coincidence of a high-fluence blazar outburst with a
  PeV-energy neutrino event}, Nature Phys. 12~(8) (2016) 807--814.
\newblock \href {http://arxiv.org/abs/1602.02012} {\path{arXiv:1602.02012}},
  \href {http://dx.doi.org/10.1038/nphys3715, 10.1038/NPHYS3715}
  {\path{doi:10.1038/nphys3715, 10.1038/NPHYS3715}}.

\bibitem{Murase:2018iyl}
K.~Murase, F.~Oikonomou, M.~Petropoulou, {Blazar Flares as an Origin of
  High-Energy Cosmic Neutrinos?}, Astrophys. J. 865~(2) (2018) 124.
\newblock \href {http://arxiv.org/abs/1807.04748} {\path{arXiv:1807.04748}},
  \href {http://dx.doi.org/10.3847/1538-4357/aada00}
  {\path{doi:10.3847/1538-4357/aada00}}.

\bibitem{Cerruti:2018tmc}
M.~Cerruti, A.~Zech, C.~Boisson, G.~Emery, S.~Inoue, J.~P. Lenain,
  {Lepto-hadronic single-zone models for the electromagnetic and neutrino
  emission of TXS 0506+056}, Mon. Not. Roy. Astron. Soc. 483 (2019) L12.
\newblock \href {http://arxiv.org/abs/1807.04335} {\path{arXiv:1807.04335}},
  \href {http://dx.doi.org/10.1093/mnrasl/sly210}
  {\path{doi:10.1093/mnrasl/sly210}}.

\bibitem{Gao:2018mnu}
S.~Gao, A.~Fedynitch, W.~Winter, M.~Pohl, {Modelling the coincident observation
  of a high-energy neutrino and a bright blazar flare}, Nat. Astron. 3~(1)
  (2019) 88--92.
\newblock \href {http://arxiv.org/abs/1807.04275} {\path{arXiv:1807.04275}},
  \href {http://dx.doi.org/10.1038/s41550-018-0610-1}
  {\path{doi:10.1038/s41550-018-0610-1}}.

\bibitem{AlvarezMuniz:2004uz}
J.~Alvarez-Muniz, P.~Meszaros, {High energy neutrinos from radio-quiet AGNs},
  Phys. Rev. D70 (2004) 123001.
\newblock \href {http://arxiv.org/abs/astro-ph/0409034}
  {\path{arXiv:astro-ph/0409034}}, \href
  {http://dx.doi.org/10.1103/PhysRevD.70.123001}
  {\path{doi:10.1103/PhysRevD.70.123001}}.

\bibitem{Farrar:2008ex}
G.~R. Farrar, A.~Gruzinov, {Giant AGN Flares and Cosmic Ray Bursts}, Astrophys.
  J. 693 (2009) 329--332.
\newblock \href {http://arxiv.org/abs/0802.1074} {\path{arXiv:0802.1074}},
  \href {http://dx.doi.org/10.1088/0004-637X/693/1/329}
  {\path{doi:10.1088/0004-637X/693/1/329}}.

\bibitem{Farrar:2014yla}
G.~R. Farrar, {Tidal disruption jets as the source of Ultra-High Energy Cosmic
  Rays}, European Physical Journal Web of Conferences 39 (2012) 07005.
\newblock \href {http://arxiv.org/abs/1411.0704} {\path{arXiv:1411.0704}}.

\bibitem{AlvesBatista:2017shr}
R.~Alves~Batista, J.~Silk, {Ultrahigh-energy cosmic rays from tidally-ignited
  white dwarfs}, Phys. Rev. D96~(10) (2017) 103003.
\newblock \href {http://arxiv.org/abs/1702.06978} {\path{arXiv:1702.06978}},
  \href {http://dx.doi.org/10.1103/PhysRevD.96.103003}
  {\path{doi:10.1103/PhysRevD.96.103003}}.

\bibitem{Zhang:2017hom}
B.~T. Zhang, K.~Murase, F.~Oikonomou, Z.~Li, {High-energy cosmic ray nuclei
  from tidal disruption events: Origin, survival, and implications}, Phys. Rev.
  D96~(6) (2017) 063007, [Addendum: Phys. Rev.D96,no.6,069902(2017)].
\newblock \href {http://arxiv.org/abs/1706.00391} {\path{arXiv:1706.00391}},
  \href {http://dx.doi.org/10.1103/PhysRevD.96.063007,
  10.1103/PhysRevD.96.069902} {\path{doi:10.1103/PhysRevD.96.063007,
  10.1103/PhysRevD.96.069902}}.

\bibitem{Wang:2015mmh}
X.-Y. Wang, R.-Y. Liu, {Tidal disruption jets of supermassive black holes as
  hidden sources of cosmic rays: explaining the IceCube TeV-PeV neutrinos},
  Phys. Rev. D93~(8) (2016) 083005.
\newblock \href {http://arxiv.org/abs/1512.08596} {\path{arXiv:1512.08596}},
  \href {http://dx.doi.org/10.1103/PhysRevD.93.083005}
  {\path{doi:10.1103/PhysRevD.93.083005}}.

\bibitem{Dai:2016gtz}
L.~Dai, K.~Fang, {Can tidal disruption events produce the IceCube neutrinos?},
  Mon. Not. Roy. Astron. Soc. 469~(2) (2017) 1354--1359.
\newblock \href {http://arxiv.org/abs/1612.00011} {\path{arXiv:1612.00011}},
  \href {http://dx.doi.org/10.1093/mnras/stx863}
  {\path{doi:10.1093/mnras/stx863}}.

\bibitem{Senno:2016bso}
N.~Senno, K.~Murase, P.~Meszaros, {High-energy Neutrino Flares from X-Ray
  Bright and Dark Tidal Disruption Events}, Astrophys. J. 838~(1) (2017) 3.
\newblock \href {http://arxiv.org/abs/1612.00918} {\path{arXiv:1612.00918}},
  \href {http://dx.doi.org/10.3847/1538-4357/aa6344}
  {\path{doi:10.3847/1538-4357/aa6344}}.

\bibitem{Lunardini:2016xwi}
C.~Lunardini, W.~Winter, {High Energy Neutrinos from the Tidal Disruption of
  Stars}, Phys. Rev. D95~(12) (2017) 123001.
\newblock \href {http://arxiv.org/abs/1612.03160} {\path{arXiv:1612.03160}},
  \href {http://dx.doi.org/10.1103/PhysRevD.95.123001}
  {\path{doi:10.1103/PhysRevD.95.123001}}.

\bibitem{Kobayashi:2004py}
S.~Kobayashi, P.~Laguna, E.~S. Phinney, P.~Meszaros, {Gravitational wave and
  x-ray signals from stellar disruption by a massive black hole}, Astrophys. J.
  615 (2004) 855--865.
\newblock \href {http://arxiv.org/abs/astro-ph/0404173}
  {\path{arXiv:astro-ph/0404173}}, \href {http://dx.doi.org/10.1086/424684}
  {\path{doi:10.1086/424684}}.

\bibitem{Haas:2012bk}
R.~Haas, R.~V. Shcherbakov, T.~Bode, P.~Laguna, {Tidal Disruptions of White
  Dwarfs from Ultra-Close Encounters with Intermediate Mass Spinning Black
  Holes}, Astrophys. J. 749 (2012) 117.
\newblock \href {http://arxiv.org/abs/1201.4389} {\path{arXiv:1201.4389}},
  \href {http://dx.doi.org/10.1088/0004-637X/749/2/117}
  {\path{doi:10.1088/0004-637X/749/2/117}}.

\bibitem{Ioka:2005er}
K.~Ioka, S.~Razzaque, S.~Kobayashi, P.~Meszaros, {TeV-PeV neutrinos from giant
  flares of magnetars and the case of SGR 1806-20}, Astrophys. J. 633 (2005)
  1013--1017.
\newblock \href {http://arxiv.org/abs/astro-ph/0503279}
  {\path{arXiv:astro-ph/0503279}}, \href {http://dx.doi.org/10.1086/466514}
  {\path{doi:10.1086/466514}}.

\bibitem{halzen}
F.~{Halzen}, H.~{Landsman}, T.~{Montaruli}, {TeV photons and Neutrinos from
  giant soft-gamma repeaters flares}, arXiv e-prints (2005)
  astro--ph/0503348\href {http://arxiv.org/abs/astro-ph/0503348}
  {\path{arXiv:astro-ph/0503348}}.

\bibitem{Ioka:2000hs}
K.~Ioka, {Magnetic deformation of magnetars for the giant flares of the soft
  gamma-ray repeaters}, Mon. Not. Roy. Astron. Soc. 327 (2001) 639.
\newblock \href {http://arxiv.org/abs/astro-ph/0009327}
  {\path{arXiv:astro-ph/0009327}}, \href
  {http://dx.doi.org/10.1046/j.1365-8711.2001.04756.x}
  {\path{doi:10.1046/j.1365-8711.2001.04756.x}}.

\bibitem{DeLaunay:2016xpf}
J.~J. DeLaunay, D.~B. Fox, K.~Murase, P.~Mészáros, A.~Keivani, C.~Messick,
  M.~A. Mostafá, F.~Oikonomou, G.~Tešić, C.~F. Turley, {Discovery of a
  transient gamma-ray counterpart to FRB 131104}, Astrophys. J. 832~(1) (2016)
  L1.
\newblock \href {http://arxiv.org/abs/1611.03139} {\path{arXiv:1611.03139}},
  \href {http://dx.doi.org/10.3847/2041-8205/832/1/L1}
  {\path{doi:10.3847/2041-8205/832/1/L1}}.

\bibitem{Murase:2016sqo}
K.~Murase, K.~Kashiyama, P.~Mészáros, {A Burst in a Wind Bubble and the
  Impact on Baryonic Ejecta: High-Energy Gamma-Ray Flashes and Afterglows from
  Fast Radio Bursts and Pulsar-Driven Supernova Remnants}, Mon. Not. Roy.
  Astron. Soc. 461~(2) (2016) 1498--1511, [erratum: Mon. Not. Roy. Astron.
  Soc.467,no.3,3542(2017)].
\newblock \href {http://arxiv.org/abs/1603.08875} {\path{arXiv:1603.08875}},
  \href {http://dx.doi.org/10.1093/mnras/stw1328, 10.1093/mnras/stx310}
  {\path{doi:10.1093/mnras/stw1328, 10.1093/mnras/stx310}}.

\bibitem{Murase:2016ysq}
K.~Murase, P.~Meszaros, D.~B. Fox, {Fast Radio Bursts with Extended Gamma-Ray
  Emission?}, Astrophys. J. 836~(1) (2017) L6.
\newblock \href {http://arxiv.org/abs/1611.03848} {\path{arXiv:1611.03848}},
  \href {http://dx.doi.org/10.3847/2041-8213/836/1/L6}
  {\path{doi:10.3847/2041-8213/836/1/L6}}.

\bibitem{pbhamon}
{Gordana Te\v{s}i\'c}, {Primordial black hole searches with AMON}, PoS ICRC2015
  236 (2015) 8.

\bibitem{icecube}
M.~Aartsen, M.~Ackermann, J.~Adams, et~al., {The IceCube realtime alert sytem},
  Astropart. Phys. 92 (2017) 30--41.

\bibitem{hawc}
A.~U. {Abeysekara}, A.~{Albert}, R.~{Alfaro}, et~al., {Observation of the Crab
  Nebula with the HAWC Gamma-Ray Observatory}, ApJ 843 (2017) 17.

\bibitem{fermi}
M.~Ackermann, M.~Ajello, A.~Albert, et~al., {The \textit{Fermi} Large Area
  Telescope on Orbit: Event Classification,Instrument Response Functions, and
  Calibration}, ApJS 203.

\bibitem{ligo}
G.~Harry, the LIGO Scientific~Collaboration, {Advance LIGO: the next generation
  of gravitational wave detectors}, Class. Quantum Grav. 27 (2010) 12.

\bibitem{virgo}
F.~Acernese, P.~Amico, M.~Alshourbagy, et~al., {The status of coalescing
  binaries search code in Virgo, and the analysis of C5 data}, Class. Quantum
  Grav. 23 (2006) 10.

\bibitem{veritas}
J.~Holder, E.~Aliu, T.~Arlen, et~al., {VERITAS: the Very Energetic Radiation
  Imaging Telescope Array System}, Astropart.Phys. 11 (1999) 345--349.

\bibitem{master}
V.~Lipunov, V.~Kornilov, E.~Gorbovskoy, et~al., {Master Robotic Net}, Advances
  in Astronomy 2010 (2009) 275--285.

\bibitem{swift}
N.~Gehrels, G.~Chincarini, P.~Giommi, et~al., {The Swift gamma-ray burst
  mission}, ApJ 611 (2004) 1005--1020.

\bibitem{antares}
M.~Ageron, J.~Aguilar, I.~A. Samarai, et~al., {ANTARES: The first undersea
  neutrino telescope}, Nucl. Instrum. Methods Phys. Res. A 656 (2011) 11--38.

\bibitem{fact}
H.~Anderhub, M.~Backes, A.~Biland, et~al., {Design and operation of FACT - the
  first G-APD Cherenkov telescope}, Journal of Instrumentation 8 (2013) 40.

\bibitem{fermigbm}
C.~Meegan, G.~Lichti, P.~Bhat, et~al., {The Fermi Gamma-ray Burst Monitor}, ApJ
  702 (2009) 791--804.

\bibitem{auger}
J.~Abraham, P.~Abreu, M.~Aglietta, et~al., {Trigger and aperture of the surface
  detector array of the Pierre Auger Observatory}, Nucl. Instrum. Methods Phys.
  Res. A 613 (2010) 29.

\bibitem{hess}
F.Aharonian, A.~Akhperjanian, A.~Bazer-Bachi, et~al., {Observations of the Crab
  Nebula with H.E.S.S }, Astron. Astrophys. 457 (2006) 899.

\bibitem{magic}
J.Cortina, F.~Goebel, T.~Schweizer, et~al., {Technical Performance of the MAGIC
  Telescopes}, ArXiv e-prints (2009) 4\href {http://arxiv.org/abs/0907.1211}
  {\path{arXiv:0907.1211}}.

\bibitem{LMT}
M.~Heyer, G.~W. Wilson, R.~Gutermuth, other, {Early science with the Large
  Millimetre Telescope: fragmentation of molecular clumps in the Galaxy},
  Monthly Notices of the Royal Astronomical Society 473 (2018) 14.

\bibitem{LCOGT}
T.~Boroson, T.~Brown, A.~Hjelstrom, et~al., {Science operations for LCOGT: a
  global telescope network}, Proc.SPIE 9149 (2014) 10.

\bibitem{ztf}
R.~Smith, D.~R., B.~C., et~al., {The Zwicky transient facility observing
  system}, Proc. SPIE 9147 914779 (2014) 1.

\bibitem{numpy}
T.~Oliphant, A guide to numpy, vol 1, Trelgol Publishing USA (2006) 371.

\bibitem{scipy}
E.~Jones, T.~Oliphant, P.~Peterson, et~al.,
  \href{http://www.scipy.org/}{{SciPy}: Open source scientific tools for
  {Python}}, [Online; accessed <today>] (2001--).
\newline\urlprefix\url{http://www.scipy.org/}

\bibitem{astropy}
T.~Robitaille, T.~E., P.~Greenfield, et~al., {Astropy: A community python
  package for Astronomy}, Astron. Astrophys. 558 (2013) 9.

\bibitem{voevent}
R.~Seaman, R.~Williams, A.~Allan, et~al.,
  \href{http://www.ivoa.net/documents/VOEvent/}{{Sky Event Reporting
  Metadata}}, International Virtual Observatory Alliance.
\newline\urlprefix\url{http://www.ivoa.net/documents/VOEvent/}

\bibitem{colinICLAT}
C.~Turley, D.~Fox, A.~Keivani, et~al., {A Coincidence Search for Cosmic
  Neutrino and Gamma-Ray Emitting Sources Using IceCube and Fermi-LAT Public
  Data}, ApJ 863 (2018) 9.

\bibitem{fisher}
R.~Fisher, {Statistical Methods for Research Workers}, 4th Edition, Oliver \&
  Boyd, 1934.

\bibitem{colinBlazar}
C.~F. {Turley}, D.~B. {Fox}, K.~{Murase}, et~al., {Search for Blazar
  Flux-correlated TeV Neutrinos in IceCube 40-string Data}, ApJ 833 (2016) 117.
\newblock \href {http://arxiv.org/abs/1608.08983} {\path{arXiv:1608.08983}},
  \href {http://dx.doi.org/10.3847/1538-4357/833/1/117}
  {\path{doi:10.3847/1538-4357/833/1/117}}.

\end{thebibliography}

\end{document}